\documentclass[aps,twocolumn,
superscriptaddress,
footinbib,
noeprint,
longbibliography,
prb]{revtex4-2}

\usepackage[utf8]{inputenc}
\usepackage{mathtools,amssymb,bm}
\usepackage{graphicx}
\usepackage{epstopdf}
\usepackage{latexsym}
\usepackage[normalem]{ulem} 
\usepackage[caption=false, position=top]{subfig}
\usepackage[usenames,dvipsnames]{color}
\usepackage{hyperref}
\usepackage{natbib}
\usepackage{xcolor}
\usepackage{physics}
\usepackage{multirow}
\usepackage{stackengine}
\usepackage{dsfont}
\usepackage{bbold}
\usepackage{nicefrac}
\usepackage[inkscapelatex=false]{svg}
\usepackage{enumitem}
\usepackage{float}
\usepackage[export]{adjustbox}

\usepackage{framed}
\graphicspath{{gfx/}}

\newcommand{\comment}[1]{}

\def\equationautorefname~#1\null{Equation~(#1)\null}
\begin{document}

\title{A metronome spin stabilizes time-crystalline dynamics}
\date{\today}

\author{Niklas Euler}
\thanks{Both authors contributed equally to this work.}
\affiliation{Physikalisches Institut, Ruprecht-Karls-Universität Heidelberg, Im Neuenheimer Feld 226, 69120 Heidelberg, Germany}
\affiliation{Institut für Festkörpertheorie und Optik, Friedrich-Schiller-Universität Jena, Max-Wien-Platz 1, 07743 Jena, Germany}
\author{Adrian Braemer}
\thanks{Both authors contributed equally to this work.}
\affiliation{Physikalisches Institut, Ruprecht-Karls-Universität Heidelberg, Im Neuenheimer Feld 226, 69120 Heidelberg, Germany}
\author{Luca Benn}
\affiliation{Physikalisches Institut, Ruprecht-Karls-Universität Heidelberg, Im Neuenheimer Feld 226, 69120 Heidelberg, Germany}
\author{Martin Gärttner}
\email{martin.gaerttner@uni-jena.de}
\affiliation{Institut für Festkörpertheorie und Optik, Friedrich-Schiller-Universität Jena, Max-Wien-Platz 1, 07743 Jena, Germany}
%\affiliation{Physikalisches Institut, Ruprecht-Karls-Universität Heidelberg, Im Neuenheimer Feld 226, 69120 Heidelberg, Germany}

\begin{abstract}
We investigate a disorder-free quantum Ising chain subject to a time-periodic drive that rotates each spin by an angle $\pi(1-\epsilon_i)$.
In case all spins experience the same deviation $\epsilon$ and the system is initialized in a fully polarized state, the dynamics is known to be time-crystalline: the magnetization of the system exhibits period-doubled oscillations for timescales that grow exponentially with the length of the chain.
In this work, we study the effect of a deviation $\epsilon$ that differs between spins. We find that reducing $\epsilon$ for a single spin drastically enhances the lifetime of spatio-temporal order, suggesting the name ``metronome" spin.
Employing perturbative arguments in an average Hamiltonian picture, we explain this observation for initial states with macroscopic bulk magnetization.
Furthermore, in the case of random bitstring initial states, we report the enhancement of the lifetime of a topological edge mode, which can also be understood in the same picture. 
Finally, we discuss an altered geometry in which the metronome spin is not directly part of the chain, affecting the dynamics in different ways in the two scenarios considered.
Our findings unveil the intricate dynamics that emerge in Floquet systems under the influence of a spatially varying drive, thereby uncovering new avenues for Floquet engineering.
\end{abstract}

\maketitle

\section{Introduction}
\label{sec:introduction}

For the longest time, stable physical phases of matter were thought to be a concept exclusive to equilibrium physics. However, with the pioneering work of Wilczek and Shapere \cite{wilczek_quantum_2012, shapere_classical_2012}, Watanabe and Oshikawa \cite{watanabe_absence_2015}, and others, it became clear that out-of-equilibrium phases of matter are not only possible but also offer features beyond equilibrium phases \cite{else_floquet_2016, else_discrete_2020}. One of the most prevalent categories of systems in which such phases have been demonstrated is Floquet setups, that is, periodically driven systems. Instead of heating up, they can display long-lived period-doubled spatio-temporal order with remarkable stability with respect to perturbations of the drive. Due to their discrete time-translation symmetry breaking, they have been dubbed ``Floquet Time Crystals" or ``Discrete Time Crystals" (DTC) and have gained significant attention among the theoretical and experimental communities over the last decade \cite{khemani_brief_2019}.
Initially, many-body localization (MBL) was considered to be the main mechanism for stabilizing long-lived dynamics \cite{abanin_theory_2016, khemani_phase_2016, yao_discrete_2017, burau_fate_2021}. However, over the years, a multitude of other processes have been shown to lead to time-crystalline behavior in different systems, including weakly broken symmetries \cite{luitz_prethermalization_2020}, prethermalization \cite{else_prethermal_2017, weidinger_floquet_2017, long_phenomenology_2023}, domain-wall confinement \cite{collura_discrete_2022}, among others \cite{ho_critical_2017, russomanno_floquet_2017, gong_discrete_2018, huang_clean_2018}. Experimentally, time-crystalline dynamics has been observed on a variety of platforms, such as nitrogen vacancy centers \cite{choi_observation_2017, choi_probing_2019, he_quasi-floquet_2023}, NMR systems \cite{rovny_31mathrmp_2018, rovny_observation_2018, autti_observation_2018}, trapped ions \cite{zhang_observation_2017}, Rydberg atoms \cite{wadenpfuhl_emergence_2023, wu_observation_2023}, and also superconducting qubits \cite{mi_time-crystalline_2022, frey_realization_2022}, to name a few.

Drives are commonly realized by periodically rotating all spins by a fixed angle. One of the most striking features of time-crystalline order is the stability with respect to such a drive. The spatio-temporal structure is present not only at isolated points in parameter space (dictated by intrinsic symmetry of the interactions) but has also been observed for drives that systematically over- or undershoot the targeted rotation angles for the entire system by up to $\epsilon \lesssim 15\%$ in every drive period \cite{zhang_observation_2017, mi_time-crystalline_2022}. Contrary to the ``naive" expectation, these errors do not accumulate and lead to rapid dephasing but are instead compensated for through the different stabilization mechanisms mentioned above. This defining characteristic of a DTC motivates the classification as an out-of-equilibrium phase of matter, as extended areas of stability can be identified with respect to the parameters of the system and drive, e.g., the interaction strength and the deviation of the driving angle \cite{khemani_phase_2016, yao_discrete_2017}. Until recently \cite{schindler_floquet_2023}, drives and perturbations have typically been considered to be spatially uniform, i.e., equal for all constituents of the system. The question of whether and how the stability of spatio-temporal order extends to regimes where parts of the system are driven at different values of $\epsilon$ remains largely unanswered. It is especially unclear whether this structure is destabilized by a small subsystem driven at much higher values of $\epsilon$, or whether a modest amount of particles driven at small $\epsilon$ can stabilize an otherwise unstable system.

To investigate the impact of the spatial dependence of the drive, we consider a disorder-free spin-1/2 chain with nearest-neighbor Ising interactions and periodic driving through numerical simulations. When initialized in a fully magnetized state, such a system's magnetization is known to exhibit period-doubled oscillations for a time growing exponentially with system size, which we will call lifetime. 
Interestingly, by reducing the rotation angle deviation $\epsilon$ for a single spin of the chain, we find a drastic enhancement of the magnetization lifetime of the entire chain, as if the single spin was acting like a metronome that keeps the other spins on beat. We employ a time-averaged effective description which allows us to explain the observed behavior with the help of symmetry arguments for the bulk of the chain.

Building on these results, we study how generic initial states behave in the presence of a metronome spin. Again, we find analogous lifetime enhancements in magnetization autocorrelators, however, only for the outermost spins stemming from the existence of a topological edge mode.
Finally, we present a system geometry in which these two mechanisms can be clearly discerned.
Our results offer new insights into how local perturbations in the chain can have a strong impact on the overall lifetime of large systems.
This opens up new possibilities in the design and implementation of extended, (meta)stable phases of matter out of equilibrium, even in systems without disorder.

Following this introduction, we first give a more detailed description of the investigated system and the numerical methods used in Sec.~\ref{sec:numerics}. The results for bulk and edge stabilization are presented in Sec.~\ref{sec:results} and subsequently discussed in Sec.~\ref{sec:conclusion}.

\section{Model and methods}
\label{sec:numerics}

\begin{figure}
    \centering
    \includegraphics[width=\columnwidth]{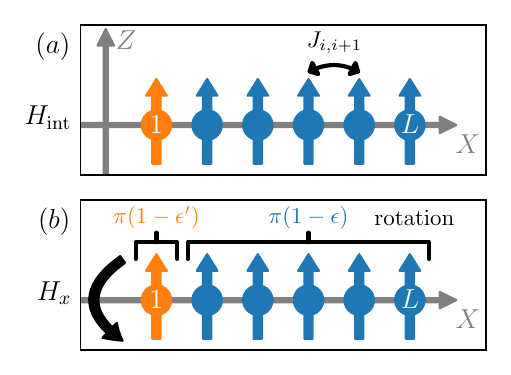}
    \caption{An illustration of the two-step Floquet cycle of the one-dimensional (1D) system considered in this work. (a) The interaction phase of the cycle given by an Ising Hamiltonian with nearest-neighbor couplings. (b) The driving phase of the Floquet cycle, realized through single-spin $s_x$ rotations. While the majority of the spins in the chain (here drawn in blue) is subjected to imperfect flips around the $x$ axis given by $\pi(1-\epsilon)$, one spin at site index $i=1$ has a differing drive-angle deviation $\epsilon_1 = \epsilon'$, resulting in $\pi(1-\epsilon')$ rotations.}
    \label{fig:visual_abstract}
\end{figure}
We study the effects of spatially nonuniform Floquet driving through numerical simulation of a spin-${1}/{2}$ chain. The Floquet sequence investigated in this work consists of two parts: in a first step, the spins interact through nearest-neighbor Ising couplings with open boundary conditions, as shown in Fig.~\ref{fig:visual_abstract}(a). Second, the spins are subjected to unitary rotations by $\pi(1-\epsilon_i)$, with $i$ indicating the site index. We consider the case where the first spin is driven with $\epsilon_{i=1}=\epsilon'$ and all other spins with $\epsilon_{i>1}=\epsilon$ [Fig.~\ref{fig:visual_abstract}(b)]. Thus, this configuration represents a uniformly driven chain with a local perturbation at one boundary site, obeying a spin-flip symmetry in the absence of $z$ fields. One cycle of this time-periodic evolution is captured by the Hamiltonian
\begin{align}
    H = \begin{cases}
        \displaystyle H_{\mathrm{int}} = \sum_{i=1}^{L-1} J_{i,i+1}s_z^is_z^{i+1} + \sum_{i=1}^{L} h_is_z^i,~ 0 \leq t \leq t_1\\
        \displaystyle H_x = \sum_{i=1}^{L}(1-\epsilon_i)s_x^i,\hspace{0.7cm} t_1 < t < t_1+\pi\eqqcolon T
    \end{cases}, \label{eq:time_dependent_hamiltonian}
\end{align}
with $s_{\{x,y,z\}}^i=\sigma_{\{x,y,z\}}^i/2$ being the single spin operators. The evolution governed by this Hamiltonian induces (imperfect) periodic flipping of the magnetization of the spin chain with period $2T$, which is twice the original period of the Hamiltonian.
Here, we are especially interested in how decreasing the deviation of the rotation angle, $\epsilon'$, for a single spin, which we call the metronome spin, affects the dynamics of spatially distant spins at late times. While we set $h_i = 0$ in the main text, we also study chains with random fields and disordered couplings in Appendix~\ref{app:disorder} and disorder-free chains with the metronome in the center, $\epsilon_{i=\frac{L+1}{2}}=\epsilon'$, in Appendix \ref{app:center}.

The stroboscopic evolution of the system, that is, evaluated only once at the beginning of every cycle, is given by the Floquet evolution operator $U_F$, which propagates the system through one cycle of the Floquet sequence. 
To gain a better understanding of the stroboscopic dynamics, one would like to find a time-independent Floquet Hamiltonian $H_F$ that generates the Floquet time evolution operator, such that
\begin{align}
    U_F = e^{-iH_x\pi}e^{-iH_{\mathrm{int}}t_1}\eqqcolon e^{-iH_FT}.\label{eq:floquet_unitary}
\end{align}
In most cases, there is no straightforward way to obtain $H_F$ analytically, but one can expand $H_F$ in the so-called Magnus series \cite{magnus_exponential_1954, blanes_magnus_2009}. By construction, $H_F$ is guaranteed to be Hermitian at all orders. We give the first two terms of the expansion,
\begin{subequations}
    \begin{align}
    H_F &= H_F^{(0)} + H_F^{(1)} + \ldots\,,\\
    H_F^{(0)} &= \frac{1}{T}\int_0^T\mathrm{d}t\,H(t),\\
    H_F^{(1)} &= \frac{1}{2Ti}\int_0^T\,\mathrm{d}t\int_0^t\,\mathrm{d}t'\,\left[H(t),H(t')\right],
    \end{align}
\end{subequations}
with the first term $H_F^{(0)}$ often being referred to as the \textit{average Hamiltonian}.

By computing the average over one period, one has to include the large $\pi(1-\epsilon_i)s_x^i$ rotations in $H_F^{(0)}$, 
\begin{equation}
\begin{split}
        H_{F,1P}^{(0)} =  \frac{1}{t_1+\pi}\Biggl[t_1\Biggl(&\sum_{i=1}^{L-1} J_{i,i+1}s_z^is_z^{i+1} + \sum_{i=1}^{L}h_i s_z^i\Biggr)\\
     +\pi&\sum_{i=1}^{L}(1-\epsilon_i)s_x^i\Biggr],\label{eq:naive_avg}
    \end{split}
\end{equation}
which is detrimental to the convergence of the Magnus series \cite{blanes_magnus_2009}. As stated earlier, the magnetization-flipping dynamics is period-doubled with respect to the time-dependent Hamiltonian. Therefore, if averaged over two periods, one not only mostly cancels the spin rotations but also averages out any random $y/z$-fields. The newly obtained two-period averaged effective Hamiltonian has the form of a transverse-field Ising model (TFIM),
\begin{align}
   H_{F,2P}^{(0)} =  \frac{1}{t_1+\pi}\left(t_1\sum_{i=1}^{L-1} J_{i,i+1}s_z^is_z^{i+1} -\pi\sum_{i=1}^{L}\epsilon_is_x^i\right),\label{eq:2period_avg}
\end{align}
hereafter only referred to as $H_{F}^{(0)}$. Alternatively, this Hamiltonian can also be derived by applying a toggling-frame transformation and subsequently taking the average over one cycle \cite{khemani_brief_2019}. This effective description retains the spin-flip symmetry present in the original time-dependent model.

\section{Results}
\label{sec:results}
In this section, we investigate numerically the lifetimes of various multi-spin and single-spin observables at stroboscopic times. We employ exact evolution according to the full Floquet unitary given in Eq.~\eqref{eq:floquet_unitary} in addition to the effective evolution with an TFIM as derived in Eq.~\eqref{eq:2period_avg} for comparison. Here, we study systems with $L=14$ spins and set $J_{ij}\eqqcolon J=1$, $h_i = 0$, $\epsilon = 0.1$, and $\epsilon' = 10^{-5}$, if not otherwise specified.

\subsection{Bulk lifetime enhancement for polarized initial states}
\label{sec:results_polarized}

\begin{figure}
    \centering
    \includegraphics[width=\columnwidth]{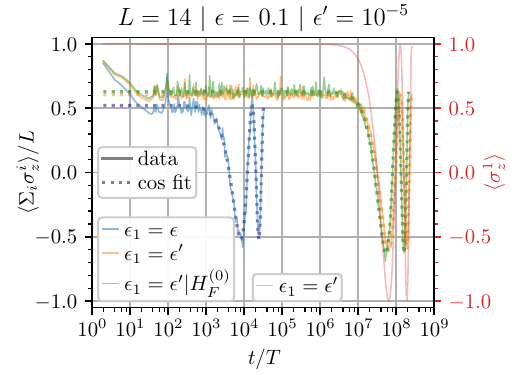}
    \caption{The global $z$ magnetization of the spin chain of length $L=14$ starting in a fully polarized stated subjected to different driving schemes. We show the exact stroboscopic dynamics of the chain at even period numbers with and without a metronome spin at one boundary site and the average Hamiltonian given in Eq.~\eqref{eq:2period_avg}. Configurations that include a metronome spin display a lifetime enhancement of several orders of magnitude. All data are well described through numerical cosine fits. On the second axis (in red font) the single-spin magnetization of the metronome spin itself is displayed. The lifetime of the magnetization of the metronome coincides with the lifetime of the total magnetization. For better visibility, only the first two oscillation cycles are plotted for each curve.}
    \label{fig:time_trace_boundary}
\end{figure}

We start by considering the dynamics of the global magnetization $\langle\sum_i\sigma_z^i\rangle/L$ for a polarized initial state $\ket{\Psi_\mathrm{init}}=\ket{\uparrow\ldots\uparrow}$, as shown on a logarithmic time axis in Fig.~\ref{fig:time_trace_boundary}. Only even period numbers are probed, so that the underlying spin-flipping dynamics is hidden in the shown simulation. The magnetization shows an initial decline that lasts $\approx10^2$ periods of the drive, largely independent of the presence of a metronome spin. Subsequently, for both with and without metronome spin we observe slow oscillations of the magnetization, which manifest themselves as extended plateaus of nonvanishing magnetization due to the log-linear axes choice in Fig.~\ref{fig:time_trace_boundary}. The macroscopic magnetization indicates that large parts of the chain retain some of its initial polarization. The duration of this plateau is strongly dependent on the angle deviation of the metronome spin drive $\epsilon'$ and, in the case of an active metronome spin, lasts $\approx10^7$ periods instead of $\approx 10^3$ periods without the metronome. The single-spin magnetization of the metronome spin, $\langle\sigma_z^1\rangle$, as shown on the right axis of Fig.~\ref{fig:time_trace_boundary}, qualitatively demonstrates the same behavior and has a lifetime similar to that of $\langle\sum_i\sigma_z^i\rangle/L$. For all data taken, the evolution under the two-period average of the Floquet Hamiltonian $H_F^{(0)}$ is in satisfactory agreement with the full Floquet evolution (the green line shows this for the case with metronome spin), indicating that it is a sufficiently good description of the full stroboscopic evolution. Therefore, we can safely focus on the simpler time-independent $H_{F}^{(0)}$ to better understand the observed behavior.

Our chosen polarized initial state is the superposition of the two lowest-energy eigenstates of $H_F^{(0)}$ which, for $\epsilon\ll J$, are well-approximated by the two parity states, \hbox{$\ket{\pm}_L = (\ket{\uparrow\ldots\uparrow} \pm \ket{\downarrow\ldots\downarrow})/\sqrt{2}$}. At finite $\epsilon$, states with domain-wall excitations are admixed (domain-wall dressing), leading to the observed initial fast decay. The energy gap between the two lowest-lying states is $\propto\epsilon^L$ in the uniform case by a perturbative argument, considering that all $L$ spins are being flipped through off-resonant coupling to excited states. Thus, the gap vanishes in the limit $L\rightarrow\infty$, making the two states degenerate. In the case of $L=14$ presented here, the gap is still finite and leads to slow Rabi oscillations of period $T_R$ between the two polarized states, which explains the observed behavior. The data show good agreement with the numerical cosine fits $\propto\cos{(2\pi t/T_R)}$, as also plotted in Fig.~\ref{fig:time_trace_boundary}, with $T_R(\epsilon_1=\epsilon)=(1.641\pm0.006)10^4 T$ and $T_R(\epsilon_1=\epsilon')=(1.281\pm0.004)10^8 T$. This difference in the length of the period of four orders of magnitude is expected in the average Hamiltonian picture, as the energy gap given above is inversely proportional to the Rabi-oscillation period $T_R\propto \epsilon^{-L}$. By endowment of one spin with reduced $\epsilon'$, one obtains
\begin{align}
    T_R\propto \epsilon^{-L+1}(\epsilon')^{-1},\label{eq:lifetime_perturbative}
\end{align}
which yields the observed difference for the values used of $\epsilon=0.1$ and $\epsilon'=10^{-5}$. 

\begin{figure}
    \centering
    \includegraphics[width=\columnwidth]{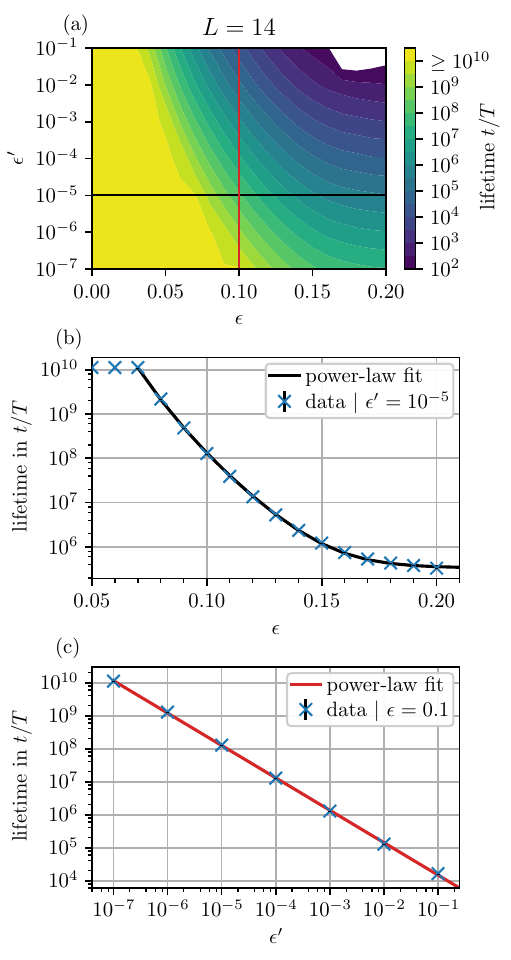}
    \caption{The lifetime of the global $z$ magnetization $\langle\sum_i \sigma_z^i\rangle/L$ for a range of drive deviation parameters $\epsilon$ and $\epsilon'$ including a metronome spin on one boundary site is shown in (a). (b) A horizontal cut through this plane at fixed $\epsilon'=10^{-5}$. (c) A vertical cut at fixed $\epsilon=0.1$, as indicated by the black and red lines in (a), respectively. All data points have been obtained through cosine fits, as shown in Fig.~\ref{fig:time_trace_boundary}. Lifetimes above $t = 10^{10}T$ and below $t = 10^{2}T$ cannot be adequately resolved and are therefore exempt from the fit.}
    \label{fig:ordered_L14_cuts}
\end{figure}

Next, we systematically investigate the dependence of the global magnetization lifetime $T_R$ on the deviations of the drive angle. For this, we repeat the procedure explained above for a number of combinations of the values of $\epsilon$ and $\epsilon'$. The results are shown in Fig.~\ref{fig:ordered_L14_cuts}(a), where we have probed a wide regime of drive parameters. At small $\epsilon$, we observe lifetimes that exceed the resolved duration of $10^{10}$ Floquet cycles (yellow region). For fixed $\epsilon=0.1$, we find that $T_R$ is approximately inversely proportional to $\epsilon'$, $T_R \propto (\epsilon')^\alpha$ with $\alpha = {-0.982\pm0.007}$, as shown in Fig.~\ref{fig:ordered_L14_cuts}(b), which is consistent with the reasoning presented in Eq.~\eqref{eq:lifetime_perturbative}. For fixed $\epsilon'=10^{-5}$, the recorded lifetimes follow a power law with offset, $T_R/T \approx a \epsilon^\beta + (3.35\pm0.09)10^5$, with $\beta = {-12.29\pm0.03}$. This value of $\beta$ roughly agrees with the expectation $\beta = {-13}$ implied by Eq.~\eqref{eq:lifetime_perturbative}. However, the observed convergence to a nonzero lifetime in the limit of large $\epsilon$ is not predicted by this perturbative picture (cf. Eq.~\ref{eq:lifetime_perturbative}).

This behavior can be understood by taking $\epsilon'\rightarrow 0$. In this limit, the dynamics of the metronome spin effectively decouple from the bulk of the chain, since the metronome cannot leave the manifold of $\{\ket{\uparrow},\ket{\downarrow}\}$, alternating between the two states in every Floquet cycle. One can now write down a Hamiltonian restricted to the bulk of the chain, where the coupling between the metronome and its neighboring spin, $s_z^1s_z^2$, can be replaced by an effective field on the second spin of the chain,
\begin{align}
    H_{F,\mathrm{bulk}}^{(0)} =  \frac{1}{T}\left(\tilde{h}s_z^{2} + t_1\sum_{i=2}^{L-1} J_{i,i+1}s_z^is_z^{i+1} -\pi\sum_{i=2}^{L}\epsilon_i s_x^i\right)\label{eq:bulk}.
\end{align}
The new field term effectively breaks the spin-flip symmetry of the original Hamiltonian $H_{F}^{(0)}$ in the bulk and thus introduces an energy gap between the two polarized states. Therefore, the prepared polarized state is no longer the superposition of the two lowest energy eigenstates but, rather, very close to the lowest eigenstate of $H_{F,\mathrm{bulk}}^{(0)}$, resulting in a stable magnetization plateau. For cases where $\epsilon'\ll1$ the metronome spin stays close to the $\{\ket{\uparrow},\ket{\downarrow}\}$ manifold for extended periods of time before it and, subsequently, the rest of the chain dephases. However, in the large $\epsilon$ limit, large parts of the chain farther away from the metronome lose their magnetization much earlier due to domain-wall excitations. Still, since the metronome is largely decoupled in its dynamic from the rest of the chain, it retains nonvanishing magnetization even at late times, keeping the magnetization plateau alive, albeit at a lower value $\langle\sum_i \sigma_z^i\rangle/L\sim\mathcal{O}(1/L)$.
This explains the observed saturation behavior in the lifetime dependence at large $\epsilon$ in Fig.~\ref{fig:ordered_L14_cuts}(b).

\subsection{Edge mode enhancement for random bit-string initial states}
\label{sec:results_edgemode}

\begin{figure}
    \centering
    \includegraphics[width=\columnwidth]{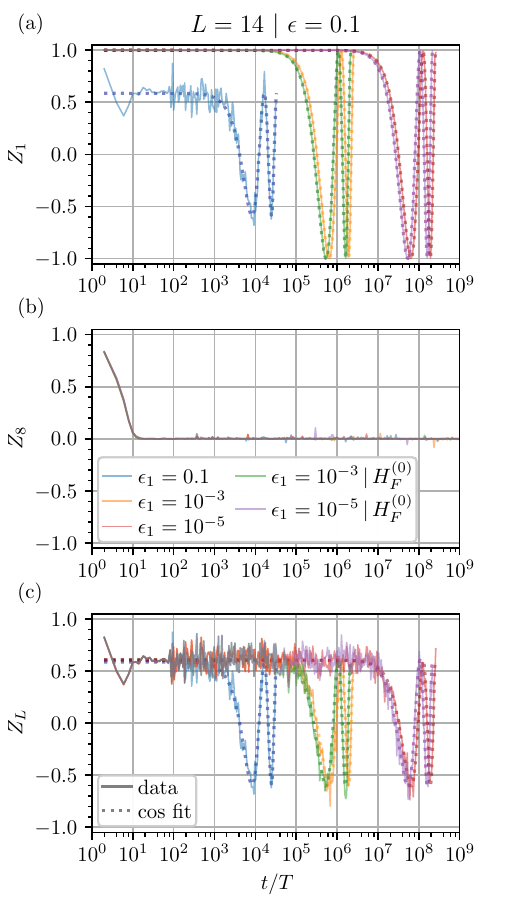}
    \caption{The $z$-magnetization autocorrelators $Z_i$ averaged over 500 random initial bitstring states for three different sites in a chain of $L=14$ spins with open boundary conditions. (a) The autocorrelator for the metronome spin at the left boundary of the chain at site index $i=1$. (b) The autocorrelator for a spin in the middle of the chain at site index $i=8$. (c) The autocorrelator for the right boundary site with $i=14$. We observe a long-lived edge mode with clear lifetime enhancement through the introduction of a metronome spin.}
    \label{fig:autocorrelators_L14}
\end{figure}

Next, we investigate how the introduction of a metronome affects the dynamics of different initial states beyond the fully polarized case.
To this end, we subject an ensemble of random bit string states, i.e., states where every spin is either $\ket{\uparrow}$ or $\ket{\downarrow}$ chosen randomly, 
to the Floquet sequence given in Eq.~\eqref{eq:time_dependent_hamiltonian}. As the magnetization of these states vanishes on average, we instead consider local magnetization autocorrelators in the rotating frame, $\langle\sigma_z^i(0)\sigma_z^i(t/T)\rangle(-1)^{\lfloor t/T\rfloor} \eqqcolon Z_i$. Three autocorrelators of selected spin sites, averaged over a set of 500 bit string initial states, are shown in Fig.~\ref{fig:autocorrelators_L14}. The three panels show the autocorrelators of the metronome spin on the site $i=1$ in Fig.~\ref{fig:autocorrelators_L14}(a), of a spin in the bulk of the chain on site $i=8$ in Fig.~\ref{fig:autocorrelators_L14}(b), and at the other chain boundary on site $i=14$ in Fig.~\ref{fig:autocorrelators_L14}(c). 

First, the results for the autocorrelator of the metronome spin itself are in line with the results for the metronome single-spin magnetization in Fig.~\ref{fig:time_trace_boundary} (right axis). The autocorrelator oscillates with full amplitude, $-1\leq Z_1\leq 1$, even at late times. Second, for sites in the bulk, we observe a rapid decline in the autocorrelator to zero, regardless of the value of $\epsilon'$. Third, we see a plateau of the autocorrelator of the opposite boundary site analogous to the dynamics of the metronome site itself, as presented in Fig.~\ref{fig:autocorrelators_L14}(c). These findings are consistent with previously reported properties of the so-called ``$0\pi$PM" out-of-equilibrium Floquet phase, originally found in disordered systems \cite{khemani_phase_2016, khemani_brief_2019}. More recently, such a phase has also been observed in a disorder-free configuration \cite{yates_long-lived_2022}. It is a symmetry-protected topological (SPT) phase \cite{kitaev_unpaired_2001, senthil_symmetry-protected_2015} with the key dynamical characteristic of long-lived period-doubled oscillations restricted to the boundaries of the system. Our data show a clear enhancement of the lifetime of the autocorrelator at the boundary sites, $Z_L$, through the introduction of the metronome spin, even though the two boundaries are separated by $L-2=12$ spins coupled only through nearest-neighbor interactions. In particular, it is not necessary to apply the stabilized drive directly to one of the two edge modes. Additional simulations of a chain with a central metronome spin reveal a similar behavior with edge-mode lifetime enhancement. More details on this additional investigation can be found in the Appendix~\ref{app:center}.

\begin{figure}
    \centering
    \includegraphics[width=\columnwidth]{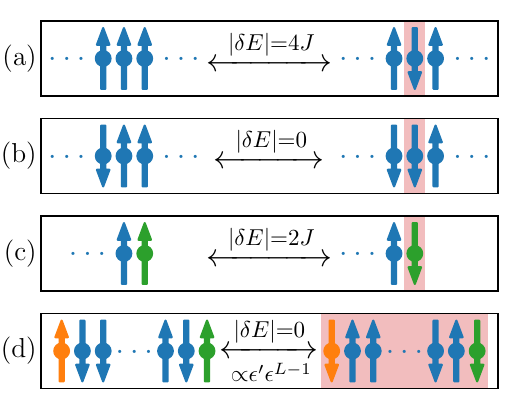}
    \caption{A visual summary of the different spin-flipping processes and their associated energy differences. The spins in the bulk are colored blue, whereas the metronome is colored orange, and the right edge spin is colored green. Flipped spins are highlighted with a red background. (a)~The creation or annihilation of two domain walls in the bulk of the chain. If the two adjacent spins are aligned, flipping the central spin results in an energy difference of $|\delta E| = 4J$. (b)~The free propagation of a domain wall. If the two adjacent spins of a central spin are antialigned, flipping the central spin is energetically degenerate, i.e., $|\delta E| = 0$. Therefore, domain walls can propagate freely along the chain, utilizing this mechanism to iteratively flip the next spin at the domain wall. (c)~The flipping of the edge spin. Flipping an edge spin always results in an energy difference of $|\delta E|=2J$, half of the bulk value, since it is coupled to only one neighboring spin. Consequently, the edge spins cannot participate in the domain-wall dynamics shown in panel (b). The resulting coupling for the first three processes is $\propto\epsilon$. (d)~The flipping of the entire chain. Flipping all spins together preserves the domain-wall structure of the chain and thus does not have an associated energy difference. This resonant process flips the edge spins at an effective rate $\propto\epsilon'\epsilon^{L-1}$.}
    \label{fig:resonant}
\end{figure}

This behavior can be understood by considering the spectral structure of the average Hamiltonian $H_{F}^{(0)}$, which we motivate in the following by a dynamical perspective. In the regime of small transverse field, the spectrum of the TFIM approximately decomposes into blocks of states with equal number of domain walls, i.e., adjacent spins pointing in opposite direction. The interaction term yields an energy difference of $2J$ per domain wall between these blocks.
The action of the field term is twofold in this view: it causes spin flips, which, in the bulk of the chain, can either create or annihilate two adjacent domain walls [see Fig.~\ref{fig:resonant}(a)] or move an existing domain wall by one site [see Fig.~\ref{fig:resonant}(b)]. The former changes the number of domain walls by $\pm 2$ and is thus off-resonant, i.e., comes at an energy cost. The latter, however, leaves the total number of domain walls invariant and thus is resonant, i.e., domain walls can propagate freely within the bulk. At the edges of the chain, any spin flip always creates or annihilates a single domain wall. This observation is at the heart of the topological protection of the edge spins: Flipping an edge spin is the only process that changes the number of domain walls by an odd amount, and thus is always off-resonant, unless both edge spins are flipped. 
One process that is always resonant and simultaneously flips both edge spins is flipping all spins [see Fig.~\ref{fig:resonant}(d)] as it corresponds to the global symmetry of the system. All other processes that alter the edge spins are strongly suppressed, because after diagonalizing the resonant domain wall dynamics in the bulk, the resulting eigenstates do not feature any other resonant transitions. This leads to the observed oscillations with frequency $\propto\epsilon'\epsilon^{L-1}$ as in the case of the fully polarized initial state.

\begin{figure}
    \centering
    \includegraphics[width=\columnwidth]{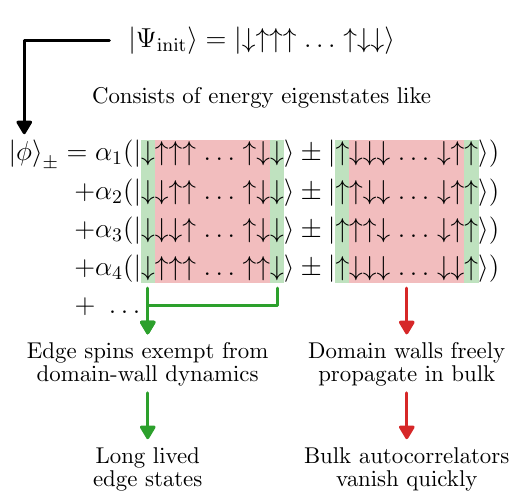}
    \caption{An illustration of the topological edge-mode protection mechanism. The system is initially prepared in a random bitstring state $\ket{\Psi_\mathrm{init}}$. Expanded in the energy eigenbasis of the average Hamiltonian $H_F^{(0)}$, the initial state has overlap with many eigenstate pairs $\ket{\phi}_\pm$ with the same number of domain walls. These eigenstate pairs each comprise a superposition of all possible domain wall placements since domain walls can move freely (highlighted in red). The dephasing between eigenstate pairs leads to the vanishing autocorrelators in the bulk. %Consequently, expectation values of autocorrelators in the bulk vanish quickly due to domain wall dephasing (highlighted in red). 
    However, since domain walls cannot propagate through the edges, edge spins (marked in green) are protected from the domain-wall dynamics. Instead, they show long coherent oscillations due to the exponentially small energy gap between the parity sectors.}
    \label{fig:flowchart}
\end{figure}

To better illustrate that last point, we translate the dynamical perspective above onto the static eigenstates of the average Hamiltonian $H_{F}^{(0)}$. Starting with the global parity symmetry, all eigenstates $\ket{\phi_\pm} \propto \ket{\phi}\pm\ket{\bar{\phi}}$ are also eigenstates of the parity operator and thus are an equal superposition of a state $\ket{\phi}$ and its spin-flipped counterpart $\ket{\bar{\phi}}$. For weak field $\epsilon\ll J$, the domain-wall number is approximately conserved, which means that each eigenstate predominantly consists of states from the same domain-wall-number sector with only minor admixtures from adjacent sectors. The observation from the dynamical viewpoint in the previous paragraph, namely that domain walls can propagate freely, here means that within the same domain-wall-number sector, the location of domain walls is ill-defined and the eigenstates are a superposition of all possible placements (see Fig.~\ref{fig:flowchart}). 

With this characterization of the eigenstates, the explanation of the observations made above is straightforward (see sketch Fig.~\ref{fig:flowchart}). Taking a bitstring initial state and expanding it in the eigenstate basis, we find it to overlap with many different eigenstates from the same sector of the domain-wall-number operator. These eigenstates dephase rapidly $\propto \mathcal{O}(\epsilon)$ and lead to the decay of autocorrelators in the bulk, as seen in Fig.~\ref{fig:autocorrelators_L14}(b). On the contrary, the edge spins can only change due to the dephasing between the parity sectors, which happens $\propto\mathcal{O}(\epsilon'\epsilon^{L-1})$. Since the splitting is identical for all components, this leads to the long coherent oscillations seen at late times in Fig.~\ref{fig:autocorrelators_L14}(a) and (c). The initial decay of the edge spin opposite to the metronome [see Fig.~\ref{fig:autocorrelators_L14}(c)] is caused by the admixture of wave-function components with a different number of domain walls.

\subsection{Adapted model with external metronome spin}
\label{sec:results_external}

\begin{figure}
    \centering
    \includegraphics[width=\columnwidth]{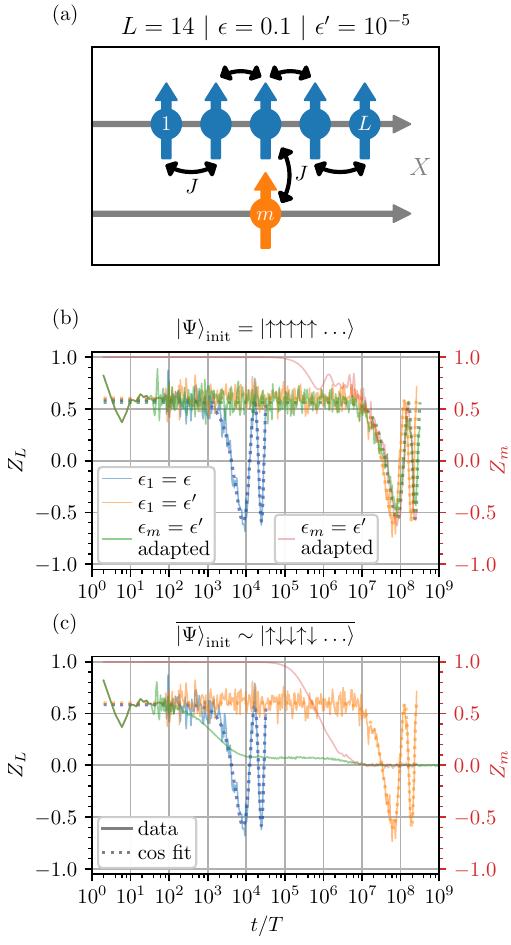}
    \caption{The setup of the adapted system with $L-1$ spins in a chain with an additional externally-coupled metronome spin attached to the center. (a) A schematic visualization of the adapted geometry. The lifetimes of the boundary-site $z$-magnetization autocorrelators $Z_L$ are shown in the next two panels for different initial states. (b) The autocorrelator of the polarized initial state. The blue and orange lines show the results for the case without metronome spin and with a metronome spin in the chain, respectively, for reference. Both configurations with a metronome display similar lifetime enhancements of the autocorrelator. (c) The autocorrelator averaged over 500 random initial bitstring states. The adapted system shows a much earlier decay of the autocorrelators compared to the configuration with the metronome spin in the chain and retains only a remnant of the original magnetization for the duration of the metronome lifetime. On the second axis in (b) and (c) (in red font) the magnetization autocorrelator $Z_m$ of the metronome spin itself in the adapted setup is displayed.}
    %\ab{Swap panels b) and c)? Same order as in text and manuscript.}
    \label{fig:autocorrelators_L14_external}
\end{figure}

To clearly separate the two described stabilization mechanisms introduced in Secs.~\ref{sec:results_polarized} and \ref{sec:results_edgemode}, we modify the geometry of the model as shown in Fig.~\ref{fig:autocorrelators_L14_external}(a).
Instead of attaching the metronome spin to one end of the chain, as previously shown in Fig.~\ref{fig:visual_abstract}, the metronome is coupled to the central spin, which itself is still coupled to its two neighbors in the chain. Thus, the two boundary spins are driven in the same way, and they are connected by a direct line of not actively stabilized spins in the bulk.
By the reasoning outlined in Sec.~\ref{sec:results_polarized}, one expects similar results for polarized initial states compared to the standard layout of Fig.~\ref{fig:visual_abstract}, as the argument relating to the effective symmetry breaking in the bulk still holds. However, the new configuration includes three ``edges" and one central spin coupled to three neighbors, one of which being the metronome spin. One important conceptual difference to the linear configuration is that here the number of domain walls in the main part of the chain is less strictly conserved. This results in a much weaker edge protection as the coupling between adjacent domain wall sectors is no longer strongly suppressed.

To test these hypotheses, we compute the $z$-magnetization autocorrelators of a boundary site, $Z_L$, and of the new metronome site, $Z_m$, with the results given in Figs.~\ref{fig:autocorrelators_L14_external}(b),(c). The observed lifetime behavior is in full agreement with the previous predictions. In the fully polarized case, we see analogous results, whereas for random bitstring states, the averaged autocorrelator of the edge spin (green curve) decreases rapidly to an intermediate plateau before vanishing completely. The timescale of the larger first decay is comparable to the lifetime of the nonstabilized chain ($t/T \approx 10^3$), and the second late-time decay coincides with the dephasing of the metronome spin. The initial decay stems from the multitude of different couplings between domain-wall sectors and the small remaining autocorrelations are protected by the spin-flip parity that is broken on timescales $\ll \epsilon'$ where the metronome is still fully polarized.

\section{Conclusion}
\label{sec:conclusion}

In this work, we have shown that near-resonant driving of a single spin can significantly increase the lifetime of long-range order in periodically driven systems. In particular, the stabilization is not based on disorder-induced MBL; instead, we have identified two distinct mechanisms that lead to long-lived bulk and edge spins, respectively. For polarized states, an argument concerning the breaking of spin-flip symmetry was found to explain the increased bulk magnetization lifetimes. Subsequent studies revealed a lifetime enhancement of stable oscillations on the boundary spins in arbitrary bitstring initial states.
We argued that the reason for the slowed edge-mode decay is that the metronome spin leads to a suppression of resonant higher-order processes.
Finally, we discussed another setup with external stabilization to the chain and thus no edge-mode enhancement to clearly highlight the two different mechanisms identified before.

Thus, our work introduces novel stabilization mechanisms suitable for ordered and, in particular, finite-size systems. The bulk-stabilization argument relies on the effective symmetry breaking introduced through one metronome spin, which is not affected by the length of the chain. Similarly, the energy offset of flipping edge spins compared to the bulk is linked to open boundary conditions, leading to enhancement of stable oscillation even for short chains. Therefore, both processes enable arbitrarily long-lived oscillations without taking the thermodynamic limit.

One potential future extension of this work is the study of two- and three-dimensional setups. The existence of MBL and thus disorder-stabilized DTC in these systems has been the subject of ongoing debate in recent years, which makes the study of alternative stabilization mechanisms an interesting direction. Moreover, the search for analogous stabilization mechanisms in other paradigmatic spin models, such as the Heisenberg XX and XXZ models, could lead to new insights into out-of-equilibrium dynamics in quantum many-body systems.

\section{Acknowledgment}
We thank Matthias Weidemüller,  Roderich Mössner, and Sebastian Geier for fruitful discussions. This work is supported by the Deutsche Forschungsgemeinschaft 
(DFG, German Research Foundation) under Germany’s Excellence Strategy 
EXC2181/1-390900948 (the Heidelberg STRUCTURES Excellence Cluster) and 
within the Collaborative Research Center SFB1225 (ISOQUANT). 
The authors acknowledge support by the state of Baden-Württemberg through bwHPC
and the DFG through Grant No INST 40/575-1 FUGG 
(Helix and JUSTUS2 clusters). For the numerical work, we used the Julia programming language \cite{bezansonJuliaFreshApproach2017}.

N.E. and A.B. contributed equally to this work.

\appendix

\section{Disorderd couplings and fields}
\label{app:disorder}
\begin{figure}
    \centering
    \includegraphics[width=\columnwidth]{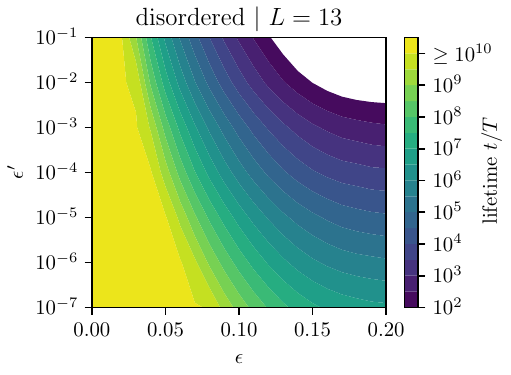}
    \caption{The disorder-averaged lifetimes of the global chain $z$ magnetization with $N=13$ spins, starting in the fully polarized initial state. The data were obtained by numerical sigmoid fits. Lifetimes above $\geq 10^{10}T$ and below $\leq 10^{2}T$ cannot be adequately resolved.}
    \label{fig:disordered_L13}
\end{figure}

\begin{figure}
    \centering
    \includegraphics[width=\columnwidth]{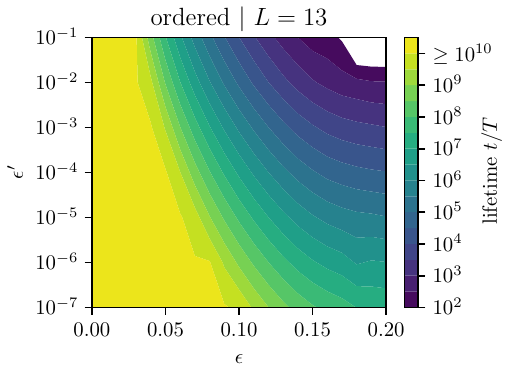}
    \caption{The lifetimes of the global ordered-chain $z$ magnetization with $N=13$ spins, starting in the fully polarized initial state. The data were obtained through numerical cosine fits. Lifetimes above $\geq 10^{10}T$ and below $\leq 10^{2}T$ cannot be adequately resolved.}
    \label{fig:ordered_L13}
\end{figure}
The stabilization mechanism proposed in Sec.~\ref{sec:results_polarized} is not based on the presence of disorder in the system. To study the interplay of disorder with the metronome, we repeat the drive-parameter scan of Fig.~\ref{fig:ordered_L14_cuts} for a disordered system.
Specifically, we subject a polarized initial state to realizations of the Hamiltonian of Eq.~\ref{eq:time_dependent_hamiltonian}, where the parameters $J_{i,i+1}$ and $h_i$ are uniform iid random variables according to $J_{i,i+1}\sim\mathcal{U}(0.5,1.5)$ and $h_{i,}\sim\mathcal{U}(-1,1)$, in analogy to simulations shown in Ref.~\cite{else_floquet_2016}. The resulting average over 250 disorder realizations of the Floquet unitary is shown in Fig.~\ref{fig:disordered_L13}. Since the increase in complexity due to the disorder average required a reduction in the size of the system to $L=13$, we also give the analogous data set for a disorder-free chain of the same length in Fig.~\ref{fig:ordered_L13}.
The averaged time traces approximately follow a sigmoid shape $\propto 1/(1+\exp{\alpha t})$, as different disorder realizations have different Rabi oscillation frequencies and cancel out at late times. The times plotted in Fig.~\ref{fig:disordered_L13} correspond to $t=1/\alpha$, so the magnetization has decreased to $\sim1/(1+e)\approx26.9\%$ of the plateau value. Comparing the two figures reveals that the behavior is qualitatively the same. However, making direct quantitative comparisons between the two data sets is not directly possible due to the differences discussed in the determination of the lifetime.

\section{Metronome spin at the center of the chain}
\label{app:center}

\begin{figure}[H]
    \centering
    \includegraphics[width=\columnwidth]{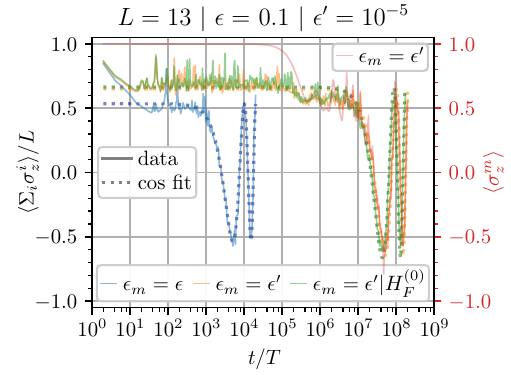}
    \caption{The global $z$ magnetization of the spin chain of length $L=13$ starting in a fully polarized state subjected to different driving schemes. We show the exact stroboscopic dynamics of the chain at even period numbers with and without a metronome spin at the central site as well as the average Hamiltonian given in Eq.~\eqref{eq:2period_avg}. The configurations that include a metronome spin display lifetime enhancements by several orders of magnitude. All data are well described through numerical cosine fits. On the second axis (in red font) we display the single-spin magnetization of the metronome spin itself. The dephasing of the magnetization of the metronome coincides with the dephasing of the plateau.}
    \label{fig:time_trace_center}
\end{figure}

Up until now, we have studied systems with the metronome attached to the end of a linear chain or to the side of it, coupled to the central spin of the chain. Now, we replace a central spin on the index $i=\lfloor(L+1)/2\rfloor\eqqcolon m$, i.e., $\epsilon_m=\epsilon'$. For odd chain lengths (here $L=13$), the spin is exactly in the middle of the chain, and the system has a spatial inversion symmetry, reducing the numerical complexity. The global $z$ magnetization of a polarized initial state is depicted in Fig.~\ref{fig:time_trace_center}.
The global magnetization of the centrally stabilized system has many similarities with that of the original setup with stabilization at the boundary, as shown in Fig.~\ref{fig:time_trace_boundary}. The system demonstrates Rabi oscillations with a similar frequency and initial magnetization amplitude. However, after $\approx 10^5$ Floquet drive cycles, the metronome-spin magnetization temporarily decays to the chain average (right axis) in Fig.~\ref{fig:time_trace_center}, before the subsequent Rabi oscillations set in.
The metronome is coupled to two neighboring spins subjected to the standard drive angle deviation, $\epsilon$, instead of the previous single spin. Therefore, the observed reduced time of the initial decay compared to the boundary metronome is consistent with this difference in chain configuration. Before that decay, the dynamics of the metronome can, in good approximation, again be considered to be largely independent of the rest of the chain. Thus, the metronome effectively decouples the two half-chains, acting as a rotating field on its two neighbors. After the decay, the two chains are coupled again, leading to the intermediate plateau before the late-time Rabi oscillations.

\bibliography{references}

%apsrev4-2.bst 2019-01-14 (MD) hand-edited version of apsrev4-1.bst
%Control: key (0)
%Control: author (8) initials jnrlst
%Control: editor formatted (1) identically to author
%Control: production of article title (0) allowed
%Control: page (0) single
%Control: year (1) truncated
%Control: production of eprint (1) enabled
\begin{thebibliography}{37}%
\makeatletter
\providecommand \@ifxundefined [1]{%
 \@ifx{#1\undefined}
}%
\providecommand \@ifnum [1]{%
 \ifnum #1\expandafter \@firstoftwo
 \else \expandafter \@secondoftwo
 \fi
}%
\providecommand \@ifx [1]{%
 \ifx #1\expandafter \@firstoftwo
 \else \expandafter \@secondoftwo
 \fi
}%
\providecommand \natexlab [1]{#1}%
\providecommand \enquote  [1]{``#1''}%
\providecommand \bibnamefont  [1]{#1}%
\providecommand \bibfnamefont [1]{#1}%
\providecommand \citenamefont [1]{#1}%
\providecommand \href@noop [0]{\@secondoftwo}%
\providecommand \href [0]{\begingroup \@sanitize@url \@href}%
\providecommand \@href[1]{\@@startlink{#1}\@@href}%
\providecommand \@@href[1]{\endgroup#1\@@endlink}%
\providecommand \@sanitize@url [0]{\catcode `\\12\catcode `\$12\catcode
  `\&12\catcode `\#12\catcode `\^12\catcode `\_12\catcode `\%12\relax}%
\providecommand \@@startlink[1]{}%
\providecommand \@@endlink[0]{}%
\providecommand \url  [0]{\begingroup\@sanitize@url \@url }%
\providecommand \@url [1]{\endgroup\@href {#1}{\urlprefix }}%
\providecommand \urlprefix  [0]{URL }%
\providecommand \Eprint [0]{\href }%
\providecommand \doibase [0]{https://doi.org/}%
\providecommand \selectlanguage [0]{\@gobble}%
\providecommand \bibinfo  [0]{\@secondoftwo}%
\providecommand \bibfield  [0]{\@secondoftwo}%
\providecommand \translation [1]{[#1]}%
\providecommand \BibitemOpen [0]{}%
\providecommand \bibitemStop [0]{}%
\providecommand \bibitemNoStop [0]{.\EOS\space}%
\providecommand \EOS [0]{\spacefactor3000\relax}%
\providecommand \BibitemShut  [1]{\csname bibitem#1\endcsname}%
\let\auto@bib@innerbib\@empty
%</preamble>
\bibitem [{\citenamefont {Wilczek}(2012)}]{wilczek_quantum_2012}%
  \BibitemOpen
  \bibfield  {author} {\bibinfo {author} {\bibfnamefont {F.}~\bibnamefont
  {Wilczek}},\ }\bibfield  {title} {\bibinfo {title} {Quantum {Time}
  {Crystals}},\ }\href {https://doi.org/10.1103/PhysRevLett.109.160401}
  {\bibfield  {journal} {\bibinfo  {journal} {Phys. Rev. Lett.}\ }\textbf
  {\bibinfo {volume} {109}},\ \bibinfo {pages} {160401} (\bibinfo {year}
  {2012})}\BibitemShut {NoStop}%
\bibitem [{\citenamefont {Shapere}\ and\ \citenamefont
  {Wilczek}(2012)}]{shapere_classical_2012}%
  \BibitemOpen
  \bibfield  {author} {\bibinfo {author} {\bibfnamefont {A.}~\bibnamefont
  {Shapere}}\ and\ \bibinfo {author} {\bibfnamefont {F.}~\bibnamefont
  {Wilczek}},\ }\bibfield  {title} {\bibinfo {title} {Classical {Time}
  {Crystals}},\ }\href {https://doi.org/10.1103/PhysRevLett.109.160402}
  {\bibfield  {journal} {\bibinfo  {journal} {Phys. Rev. Lett.}\ }\textbf
  {\bibinfo {volume} {109}},\ \bibinfo {pages} {160402} (\bibinfo {year}
  {2012})}\BibitemShut {NoStop}%
\bibitem [{\citenamefont {Watanabe}\ and\ \citenamefont
  {Oshikawa}(2015)}]{watanabe_absence_2015}%
  \BibitemOpen
  \bibfield  {author} {\bibinfo {author} {\bibfnamefont {H.}~\bibnamefont
  {Watanabe}}\ and\ \bibinfo {author} {\bibfnamefont {M.}~\bibnamefont
  {Oshikawa}},\ }\bibfield  {title} {\bibinfo {title} {Absence of {Quantum}
  {Time} {Crystals}},\ }\href {https://doi.org/10.1103/PhysRevLett.114.251603}
  {\bibfield  {journal} {\bibinfo  {journal} {Phys. Rev. Lett.}\ }\textbf
  {\bibinfo {volume} {114}},\ \bibinfo {pages} {251603} (\bibinfo {year}
  {2015})}\BibitemShut {NoStop}%
\bibitem [{\citenamefont {Else}\ \emph {et~al.}(2016)\citenamefont {Else},
  \citenamefont {Bauer},\ and\ \citenamefont {Nayak}}]{else_floquet_2016}%
  \BibitemOpen
  \bibfield  {author} {\bibinfo {author} {\bibfnamefont {D.~V.}\ \bibnamefont
  {Else}}, \bibinfo {author} {\bibfnamefont {B.}~\bibnamefont {Bauer}},\ and\
  \bibinfo {author} {\bibfnamefont {C.}~\bibnamefont {Nayak}},\ }\bibfield
  {title} {\bibinfo {title} {Floquet {Time} {Crystals}},\ }\href
  {https://doi.org/10.1103/PhysRevLett.117.090402} {\bibfield  {journal}
  {\bibinfo  {journal} {Phys. Rev. Lett.}\ }\textbf {\bibinfo {volume} {117}},\
  \bibinfo {pages} {090402} (\bibinfo {year} {2016})}\BibitemShut {NoStop}%
\bibitem [{\citenamefont {Else}\ \emph {et~al.}(2020)\citenamefont {Else},
  \citenamefont {Monroe}, \citenamefont {Nayak},\ and\ \citenamefont
  {Yao}}]{else_discrete_2020}%
  \BibitemOpen
  \bibfield  {author} {\bibinfo {author} {\bibfnamefont {D.~V.}\ \bibnamefont
  {Else}}, \bibinfo {author} {\bibfnamefont {C.}~\bibnamefont {Monroe}},
  \bibinfo {author} {\bibfnamefont {C.}~\bibnamefont {Nayak}},\ and\ \bibinfo
  {author} {\bibfnamefont {N.~Y.}\ \bibnamefont {Yao}},\ }\bibfield  {title}
  {\bibinfo {title} {Discrete {Time} {Crystals}},\ }\href
  {https://doi.org/10.1146/annurev-conmatphys-031119-050658} {\bibfield
  {journal} {\bibinfo  {journal} {Annu. Rev. Condens. Matter Phys.}\ }\textbf
  {\bibinfo {volume} {11}},\ \bibinfo {pages} {467} (\bibinfo {year}
  {2020})}\BibitemShut {NoStop}%
\bibitem [{\citenamefont {Khemani}\ \emph {et~al.}(2019)\citenamefont
  {Khemani}, \citenamefont {Moessner},\ and\ \citenamefont
  {Sondhi}}]{khemani_brief_2019}%
  \BibitemOpen
  \bibfield  {author} {\bibinfo {author} {\bibfnamefont {V.}~\bibnamefont
  {Khemani}}, \bibinfo {author} {\bibfnamefont {R.}~\bibnamefont {Moessner}},\
  and\ \bibinfo {author} {\bibfnamefont {S.~L.}\ \bibnamefont {Sondhi}},\
  }\href@noop {} {\bibinfo {title} {A brief history of time crystals}}
  (\bibinfo {year} {2019}),\ \Eprint {https://arxiv.org/abs/1910.10745}
  {arXiv:1910.10745 [cond-mat.str-el]} \BibitemShut {NoStop}%
\bibitem [{\citenamefont {Abanin}\ \emph {et~al.}(2016)\citenamefont {Abanin},
  \citenamefont {Roeck},\ and\ \citenamefont {Huveneers}}]{abanin_theory_2016}%
  \BibitemOpen
  \bibfield  {author} {\bibinfo {author} {\bibfnamefont {D.~A.}\ \bibnamefont
  {Abanin}}, \bibinfo {author} {\bibfnamefont {W.~D.}\ \bibnamefont {Roeck}},\
  and\ \bibinfo {author} {\bibfnamefont {F.}~\bibnamefont {Huveneers}},\
  }\bibfield  {title} {\bibinfo {title} {Theory of many-body localization in
  periodically driven systems},\ }\href
  {https://doi.org/10.1016/j.aop.2016.03.010} {\bibfield  {journal} {\bibinfo
  {journal} {Ann. Phys. (N. Y.)}\ }\textbf {\bibinfo {volume} {372}},\ \bibinfo
  {pages} {1} (\bibinfo {year} {2016})}\BibitemShut {NoStop}%
\bibitem [{\citenamefont {Khemani}\ \emph {et~al.}(2016)\citenamefont
  {Khemani}, \citenamefont {Lazarides}, \citenamefont {Moessner},\ and\
  \citenamefont {Sondhi}}]{khemani_phase_2016}%
  \BibitemOpen
  \bibfield  {author} {\bibinfo {author} {\bibfnamefont {V.}~\bibnamefont
  {Khemani}}, \bibinfo {author} {\bibfnamefont {A.}~\bibnamefont {Lazarides}},
  \bibinfo {author} {\bibfnamefont {R.}~\bibnamefont {Moessner}},\ and\
  \bibinfo {author} {\bibfnamefont {S.~L.}\ \bibnamefont {Sondhi}},\ }\bibfield
   {title} {\bibinfo {title} {Phase {Structure} of {Driven} {Quantum}
  {Systems}},\ }\href {https://doi.org/10.1103/PhysRevLett.116.250401}
  {\bibfield  {journal} {\bibinfo  {journal} {Phys. Rev. Lett.}\ }\textbf
  {\bibinfo {volume} {116}},\ \bibinfo {pages} {250401} (\bibinfo {year}
  {2016})}\BibitemShut {NoStop}%
\bibitem [{\citenamefont {Yao}\ \emph {et~al.}(2017)\citenamefont {Yao},
  \citenamefont {Potter}, \citenamefont {Potirniche},\ and\ \citenamefont
  {Vishwanath}}]{yao_discrete_2017}%
  \BibitemOpen
  \bibfield  {author} {\bibinfo {author} {\bibfnamefont {N.~Y.}\ \bibnamefont
  {Yao}}, \bibinfo {author} {\bibfnamefont {A.~C.}\ \bibnamefont {Potter}},
  \bibinfo {author} {\bibfnamefont {I.-D.}\ \bibnamefont {Potirniche}},\ and\
  \bibinfo {author} {\bibfnamefont {A.}~\bibnamefont {Vishwanath}},\ }\bibfield
   {title} {\bibinfo {title} {Discrete {Time} {Crystals}: {Rigidity},
  {Criticality}, and {Realizations}},\ }\href
  {https://doi.org/10.1103/PhysRevLett.118.030401} {\bibfield  {journal}
  {\bibinfo  {journal} {Phys. Rev. Lett.}\ }\textbf {\bibinfo {volume} {118}},\
  \bibinfo {pages} {030401} (\bibinfo {year} {2017})}\BibitemShut {NoStop}%
\bibitem [{\citenamefont {Burau}\ \emph {et~al.}(2021)\citenamefont {Burau},
  \citenamefont {Heyl},\ and\ \citenamefont {De~Tomasi}}]{burau_fate_2021}%
  \BibitemOpen
  \bibfield  {author} {\bibinfo {author} {\bibfnamefont {H.}~\bibnamefont
  {Burau}}, \bibinfo {author} {\bibfnamefont {M.}~\bibnamefont {Heyl}},\ and\
  \bibinfo {author} {\bibfnamefont {G.}~\bibnamefont {De~Tomasi}},\ }\bibfield
  {title} {\bibinfo {title} {Fate of algebraic many-body localization under
  driving},\ }\href {https://doi.org/10.1103/PhysRevB.104.224201} {\bibfield
  {journal} {\bibinfo  {journal} {Phys. Rev. B}\ }\textbf {\bibinfo {volume}
  {104}},\ \bibinfo {pages} {224201} (\bibinfo {year} {2021})}\BibitemShut
  {NoStop}%
\bibitem [{\citenamefont {Luitz}\ \emph {et~al.}(2020)\citenamefont {Luitz},
  \citenamefont {Moessner}, \citenamefont {Sondhi},\ and\ \citenamefont
  {Khemani}}]{luitz_prethermalization_2020}%
  \BibitemOpen
  \bibfield  {author} {\bibinfo {author} {\bibfnamefont {D.~J.}\ \bibnamefont
  {Luitz}}, \bibinfo {author} {\bibfnamefont {R.}~\bibnamefont {Moessner}},
  \bibinfo {author} {\bibfnamefont {S.}~\bibnamefont {Sondhi}},\ and\ \bibinfo
  {author} {\bibfnamefont {V.}~\bibnamefont {Khemani}},\ }\bibfield  {title}
  {\bibinfo {title} {Prethermalization without {Temperature}},\ }\href
  {https://doi.org/10.1103/PhysRevX.10.021046} {\bibfield  {journal} {\bibinfo
  {journal} {Phys. Rev. X.}\ }\textbf {\bibinfo {volume} {10}},\ \bibinfo
  {pages} {021046} (\bibinfo {year} {2020})}\BibitemShut {NoStop}%
\bibitem [{\citenamefont {Else}\ \emph {et~al.}(2017)\citenamefont {Else},
  \citenamefont {Bauer},\ and\ \citenamefont {Nayak}}]{else_prethermal_2017}%
  \BibitemOpen
  \bibfield  {author} {\bibinfo {author} {\bibfnamefont {D.~V.}\ \bibnamefont
  {Else}}, \bibinfo {author} {\bibfnamefont {B.}~\bibnamefont {Bauer}},\ and\
  \bibinfo {author} {\bibfnamefont {C.}~\bibnamefont {Nayak}},\ }\bibfield
  {title} {\bibinfo {title} {Prethermal {Phases} of {Matter} {Protected} by
  {Time}-{Translation} {Symmetry}},\ }\href
  {https://doi.org/10.1103/PhysRevX.7.011026} {\bibfield  {journal} {\bibinfo
  {journal} {Phys. Rev. X.}\ }\textbf {\bibinfo {volume} {7}},\ \bibinfo
  {pages} {011026} (\bibinfo {year} {2017})}\BibitemShut {NoStop}%
\bibitem [{\citenamefont {Weidinger}\ and\ \citenamefont
  {Knap}(2017)}]{weidinger_floquet_2017}%
  \BibitemOpen
  \bibfield  {author} {\bibinfo {author} {\bibfnamefont {S.~A.}\ \bibnamefont
  {Weidinger}}\ and\ \bibinfo {author} {\bibfnamefont {M.}~\bibnamefont
  {Knap}},\ }\bibfield  {title} {\bibinfo {title} {Floquet prethermalization
  and regimes of heating in a periodically driven, interacting quantum
  system},\ }\href {https://doi.org/10.1038/srep45382} {\bibfield  {journal}
  {\bibinfo  {journal} {Sci. Rep.}\ }\textbf {\bibinfo {volume} {7}},\ \bibinfo
  {pages} {45382} (\bibinfo {year} {2017})}\BibitemShut {NoStop}%
\bibitem [{\citenamefont {Long}\ \emph {et~al.}(2023)\citenamefont {Long},
  \citenamefont {Crowley}, \citenamefont {Khemani},\ and\ \citenamefont
  {Chandran}}]{long_phenomenology_2023}%
  \BibitemOpen
  \bibfield  {author} {\bibinfo {author} {\bibfnamefont {D.~M.}\ \bibnamefont
  {Long}}, \bibinfo {author} {\bibfnamefont {P.~J.~D.}\ \bibnamefont
  {Crowley}}, \bibinfo {author} {\bibfnamefont {V.}~\bibnamefont {Khemani}},\
  and\ \bibinfo {author} {\bibfnamefont {A.}~\bibnamefont {Chandran}},\
  }\bibfield  {title} {\bibinfo {title} {Phenomenology of the {Prethermal}
  {Many}-{Body} {Localized} {Regime}},\ }\href
  {https://doi.org/10.1103/PhysRevLett.131.106301} {\bibfield  {journal}
  {\bibinfo  {journal} {Phys. Rev. Lett.}\ }\textbf {\bibinfo {volume} {131}},\
  \bibinfo {pages} {106301} (\bibinfo {year} {2023})}\BibitemShut {NoStop}%
\bibitem [{\citenamefont {Collura}\ \emph {et~al.}(2022)\citenamefont
  {Collura}, \citenamefont {De~Luca}, \citenamefont {Rossini},\ and\
  \citenamefont {Lerose}}]{collura_discrete_2022}%
  \BibitemOpen
  \bibfield  {author} {\bibinfo {author} {\bibfnamefont {M.}~\bibnamefont
  {Collura}}, \bibinfo {author} {\bibfnamefont {A.}~\bibnamefont {De~Luca}},
  \bibinfo {author} {\bibfnamefont {D.}~\bibnamefont {Rossini}},\ and\ \bibinfo
  {author} {\bibfnamefont {A.}~\bibnamefont {Lerose}},\ }\bibfield  {title}
  {\bibinfo {title} {Discrete {Time}-{Crystalline} {Response} {Stabilized} by
  {Domain}-{Wall} {Confinement}},\ }\href
  {https://doi.org/10.1103/PhysRevX.12.031037} {\bibfield  {journal} {\bibinfo
  {journal} {Phys. Rev. X.}\ }\textbf {\bibinfo {volume} {12}},\ \bibinfo
  {pages} {031037} (\bibinfo {year} {2022})}\BibitemShut {NoStop}%
\bibitem [{\citenamefont {Ho}\ \emph {et~al.}(2017)\citenamefont {Ho},
  \citenamefont {Choi}, \citenamefont {Lukin},\ and\ \citenamefont
  {Abanin}}]{ho_critical_2017}%
  \BibitemOpen
  \bibfield  {author} {\bibinfo {author} {\bibfnamefont {W.~W.}\ \bibnamefont
  {Ho}}, \bibinfo {author} {\bibfnamefont {S.}~\bibnamefont {Choi}}, \bibinfo
  {author} {\bibfnamefont {M.}~\bibnamefont {Lukin}},\ and\ \bibinfo {author}
  {\bibfnamefont {D.}~\bibnamefont {Abanin}},\ }\bibfield  {title} {\bibinfo
  {title} {Critical {Time} {Crystals} in {Dipolar} {Systems}},\ }\href
  {https://doi.org/10.1103/PhysRevLett.119.010602} {\bibfield  {journal}
  {\bibinfo  {journal} {Phys. Rev. Lett.}\ }\textbf {\bibinfo {volume} {119}},\
  \bibinfo {pages} {010602} (\bibinfo {year} {2017})}\BibitemShut {NoStop}%
\bibitem [{\citenamefont {Russomanno}\ \emph {et~al.}(2017)\citenamefont
  {Russomanno}, \citenamefont {Iemini}, \citenamefont {Dalmonte},\ and\
  \citenamefont {Fazio}}]{russomanno_floquet_2017}%
  \BibitemOpen
  \bibfield  {author} {\bibinfo {author} {\bibfnamefont {A.}~\bibnamefont
  {Russomanno}}, \bibinfo {author} {\bibfnamefont {F.}~\bibnamefont {Iemini}},
  \bibinfo {author} {\bibfnamefont {M.}~\bibnamefont {Dalmonte}},\ and\
  \bibinfo {author} {\bibfnamefont {R.}~\bibnamefont {Fazio}},\ }\bibfield
  {title} {\bibinfo {title} {Floquet time crystal in the
  {Lipkin}-{Meshkov}-{Glick} model},\ }\href
  {https://doi.org/10.1103/PhysRevB.95.214307} {\bibfield  {journal} {\bibinfo
  {journal} {Phys. Rev. B}\ }\textbf {\bibinfo {volume} {95}},\ \bibinfo
  {pages} {214307} (\bibinfo {year} {2017})}\BibitemShut {NoStop}%
\bibitem [{\citenamefont {Gong}\ \emph {et~al.}(2018)\citenamefont {Gong},
  \citenamefont {Hamazaki},\ and\ \citenamefont {Ueda}}]{gong_discrete_2018}%
  \BibitemOpen
  \bibfield  {author} {\bibinfo {author} {\bibfnamefont {Z.}~\bibnamefont
  {Gong}}, \bibinfo {author} {\bibfnamefont {R.}~\bibnamefont {Hamazaki}},\
  and\ \bibinfo {author} {\bibfnamefont {M.}~\bibnamefont {Ueda}},\ }\bibfield
  {title} {\bibinfo {title} {Discrete {Time}-{Crystalline} {Order} in {Cavity}
  and {Circuit} {QED} {Systems}},\ }\href
  {https://doi.org/10.1103/PhysRevLett.120.040404} {\bibfield  {journal}
  {\bibinfo  {journal} {Phys. Rev. Lett.}\ }\textbf {\bibinfo {volume} {120}},\
  \bibinfo {pages} {040404} (\bibinfo {year} {2018})}\BibitemShut {NoStop}%
\bibitem [{\citenamefont {Huang}\ \emph {et~al.}(2018)\citenamefont {Huang},
  \citenamefont {Wu},\ and\ \citenamefont {Liu}}]{huang_clean_2018}%
  \BibitemOpen
  \bibfield  {author} {\bibinfo {author} {\bibfnamefont {B.}~\bibnamefont
  {Huang}}, \bibinfo {author} {\bibfnamefont {Y.-H.}\ \bibnamefont {Wu}},\ and\
  \bibinfo {author} {\bibfnamefont {W.~V.}\ \bibnamefont {Liu}},\ }\bibfield
  {title} {\bibinfo {title} {Clean {Floquet} {Time} {Crystals}: {Models} and
  {Realizations} in {Cold} {Atoms}},\ }\href
  {https://doi.org/10.1103/PhysRevLett.120.110603} {\bibfield  {journal}
  {\bibinfo  {journal} {Phys. Rev. Lett.}\ }\textbf {\bibinfo {volume} {120}},\
  \bibinfo {pages} {110603} (\bibinfo {year} {2018})}\BibitemShut {NoStop}%
\bibitem [{\citenamefont {Choi}\ \emph {et~al.}(2017)\citenamefont {Choi},
  \citenamefont {Choi}, \citenamefont {Landig}, \citenamefont {Kucsko},
  \citenamefont {Zhou}, \citenamefont {Isoya}, \citenamefont {Jelezko},
  \citenamefont {Onoda}, \citenamefont {Sumiya}, \citenamefont {Khemani},
  \citenamefont {von Keyserlingk}, \citenamefont {Yao}, \citenamefont
  {Demler},\ and\ \citenamefont {Lukin}}]{choi_observation_2017}%
  \BibitemOpen
  \bibfield  {author} {\bibinfo {author} {\bibfnamefont {S.}~\bibnamefont
  {Choi}}, \bibinfo {author} {\bibfnamefont {J.}~\bibnamefont {Choi}}, \bibinfo
  {author} {\bibfnamefont {R.}~\bibnamefont {Landig}}, \bibinfo {author}
  {\bibfnamefont {G.}~\bibnamefont {Kucsko}}, \bibinfo {author} {\bibfnamefont
  {H.}~\bibnamefont {Zhou}}, \bibinfo {author} {\bibfnamefont {J.}~\bibnamefont
  {Isoya}}, \bibinfo {author} {\bibfnamefont {F.}~\bibnamefont {Jelezko}},
  \bibinfo {author} {\bibfnamefont {S.}~\bibnamefont {Onoda}}, \bibinfo
  {author} {\bibfnamefont {H.}~\bibnamefont {Sumiya}}, \bibinfo {author}
  {\bibfnamefont {V.}~\bibnamefont {Khemani}}, \bibinfo {author} {\bibfnamefont
  {C.}~\bibnamefont {von Keyserlingk}}, \bibinfo {author} {\bibfnamefont
  {N.~Y.}\ \bibnamefont {Yao}}, \bibinfo {author} {\bibfnamefont
  {E.}~\bibnamefont {Demler}},\ and\ \bibinfo {author} {\bibfnamefont {M.~D.}\
  \bibnamefont {Lukin}},\ }\bibfield  {title} {\bibinfo {title} {Observation of
  discrete time-crystalline order in a disordered dipolar many-body system},\
  }\href {https://doi.org/10.1038/nature21426} {\bibfield  {journal} {\bibinfo
  {journal} {Nature}\ }\textbf {\bibinfo {volume} {543}},\ \bibinfo {pages}
  {221} (\bibinfo {year} {2017})}\BibitemShut {NoStop}%
\bibitem [{\citenamefont {Choi}\ \emph {et~al.}(2019)\citenamefont {Choi},
  \citenamefont {Zhou}, \citenamefont {Choi}, \citenamefont {Landig},
  \citenamefont {Ho}, \citenamefont {Isoya}, \citenamefont {Jelezko},
  \citenamefont {Onoda}, \citenamefont {Sumiya}, \citenamefont {Abanin},\ and\
  \citenamefont {Lukin}}]{choi_probing_2019}%
  \BibitemOpen
  \bibfield  {author} {\bibinfo {author} {\bibfnamefont {J.}~\bibnamefont
  {Choi}}, \bibinfo {author} {\bibfnamefont {H.}~\bibnamefont {Zhou}}, \bibinfo
  {author} {\bibfnamefont {S.}~\bibnamefont {Choi}}, \bibinfo {author}
  {\bibfnamefont {R.}~\bibnamefont {Landig}}, \bibinfo {author} {\bibfnamefont
  {W.~W.}\ \bibnamefont {Ho}}, \bibinfo {author} {\bibfnamefont
  {J.}~\bibnamefont {Isoya}}, \bibinfo {author} {\bibfnamefont
  {F.}~\bibnamefont {Jelezko}}, \bibinfo {author} {\bibfnamefont
  {S.}~\bibnamefont {Onoda}}, \bibinfo {author} {\bibfnamefont
  {H.}~\bibnamefont {Sumiya}}, \bibinfo {author} {\bibfnamefont {D.~A.}\
  \bibnamefont {Abanin}},\ and\ \bibinfo {author} {\bibfnamefont {M.~D.}\
  \bibnamefont {Lukin}},\ }\bibfield  {title} {\bibinfo {title} {Probing
  {Quantum} {Thermalization} of a {Disordered} {Dipolar} {Spin} {Ensemble} with
  {Discrete} {Time}-{Crystalline} {Order}},\ }\href
  {https://doi.org/10.1103/PhysRevLett.122.043603} {\bibfield  {journal}
  {\bibinfo  {journal} {Phys. Rev. Lett.}\ }\textbf {\bibinfo {volume} {122}},\
  \bibinfo {pages} {043603} (\bibinfo {year} {2019})}\BibitemShut {NoStop}%
\bibitem [{\citenamefont {He}\ \emph {et~al.}(2023)\citenamefont {He},
  \citenamefont {Ye}, \citenamefont {Gong}, \citenamefont {Liu}, \citenamefont
  {Murch}, \citenamefont {Yao},\ and\ \citenamefont
  {Zu}}]{he_quasi-floquet_2023}%
  \BibitemOpen
  \bibfield  {author} {\bibinfo {author} {\bibfnamefont {G.}~\bibnamefont
  {He}}, \bibinfo {author} {\bibfnamefont {B.}~\bibnamefont {Ye}}, \bibinfo
  {author} {\bibfnamefont {R.}~\bibnamefont {Gong}}, \bibinfo {author}
  {\bibfnamefont {Z.}~\bibnamefont {Liu}}, \bibinfo {author} {\bibfnamefont
  {K.~W.}\ \bibnamefont {Murch}}, \bibinfo {author} {\bibfnamefont {N.~Y.}\
  \bibnamefont {Yao}},\ and\ \bibinfo {author} {\bibfnamefont {C.}~\bibnamefont
  {Zu}},\ }\bibfield  {title} {\bibinfo {title} {Quasi-{Floquet}
  {Prethermalization} in a {Disordered} {Dipolar} {Spin} {Ensemble} in
  {Diamond}},\ }\href {https://doi.org/10.1103/PhysRevLett.131.130401}
  {\bibfield  {journal} {\bibinfo  {journal} {Phys. Rev. Lett.}\ }\textbf
  {\bibinfo {volume} {131}},\ \bibinfo {pages} {130401} (\bibinfo {year}
  {2023})}\BibitemShut {NoStop}%
\bibitem [{\citenamefont {Rovny}\ \emph
  {et~al.}(2018{\natexlab{a}})\citenamefont {Rovny}, \citenamefont {Blum},\
  and\ \citenamefont {Barrett}}]{rovny_31mathrmp_2018}%
  \BibitemOpen
  \bibfield  {author} {\bibinfo {author} {\bibfnamefont {J.}~\bibnamefont
  {Rovny}}, \bibinfo {author} {\bibfnamefont {R.~L.}\ \bibnamefont {Blum}},\
  and\ \bibinfo {author} {\bibfnamefont {S.~E.}\ \bibnamefont {Barrett}},\
  }\bibfield  {title} {\bibinfo {title} {$^{31}\mathrm{P}$ {NMR} study of
  discrete time-crystalline signatures in an ordered crystal of ammonium
  dihydrogen phosphate},\ }\href {https://doi.org/10.1103/PhysRevB.97.184301}
  {\bibfield  {journal} {\bibinfo  {journal} {Phys. Rev. B}\ }\textbf {\bibinfo
  {volume} {97}},\ \bibinfo {pages} {184301} (\bibinfo {year}
  {2018}{\natexlab{a}})}\BibitemShut {NoStop}%
\bibitem [{\citenamefont {Rovny}\ \emph
  {et~al.}(2018{\natexlab{b}})\citenamefont {Rovny}, \citenamefont {Blum},\
  and\ \citenamefont {Barrett}}]{rovny_observation_2018}%
  \BibitemOpen
  \bibfield  {author} {\bibinfo {author} {\bibfnamefont {J.}~\bibnamefont
  {Rovny}}, \bibinfo {author} {\bibfnamefont {R.~L.}\ \bibnamefont {Blum}},\
  and\ \bibinfo {author} {\bibfnamefont {S.~E.}\ \bibnamefont {Barrett}},\
  }\bibfield  {title} {\bibinfo {title} {Observation of
  {Discrete}-{Time}-{Crystal} {Signatures} in an {Ordered} {Dipolar}
  {Many}-{Body} {System}},\ }\href
  {https://doi.org/10.1103/PhysRevLett.120.180603} {\bibfield  {journal}
  {\bibinfo  {journal} {Phys. Rev. Lett.}\ }\textbf {\bibinfo {volume} {120}},\
  \bibinfo {pages} {180603} (\bibinfo {year} {2018}{\natexlab{b}})}\BibitemShut
  {NoStop}%
\bibitem [{\citenamefont {Autti}\ \emph {et~al.}(2018)\citenamefont {Autti},
  \citenamefont {Eltsov},\ and\ \citenamefont
  {Volovik}}]{autti_observation_2018}%
  \BibitemOpen
  \bibfield  {author} {\bibinfo {author} {\bibfnamefont {S.}~\bibnamefont
  {Autti}}, \bibinfo {author} {\bibfnamefont {V.~B.}\ \bibnamefont {Eltsov}},\
  and\ \bibinfo {author} {\bibfnamefont {G.~E.}\ \bibnamefont {Volovik}},\
  }\bibfield  {title} {\bibinfo {title} {Observation of a {Time} {Quasicrystal}
  and {Its} {Transition} to a {Superfluid} {Time} {Crystal}},\ }\href
  {https://doi.org/10.1103/PhysRevLett.120.215301} {\bibfield  {journal}
  {\bibinfo  {journal} {Phys. Rev. Lett.}\ }\textbf {\bibinfo {volume} {120}},\
  \bibinfo {pages} {215301} (\bibinfo {year} {2018})}\BibitemShut {NoStop}%
\bibitem [{\citenamefont {Zhang}\ \emph {et~al.}(2017)\citenamefont {Zhang},
  \citenamefont {Hess}, \citenamefont {Kyprianidis}, \citenamefont {Becker},
  \citenamefont {Lee}, \citenamefont {Smith}, \citenamefont {Pagano},
  \citenamefont {Potirniche}, \citenamefont {Potter}, \citenamefont
  {Vishwanath}, \citenamefont {Yao},\ and\ \citenamefont
  {Monroe}}]{zhang_observation_2017}%
  \BibitemOpen
  \bibfield  {author} {\bibinfo {author} {\bibfnamefont {J.}~\bibnamefont
  {Zhang}}, \bibinfo {author} {\bibfnamefont {P.~W.}\ \bibnamefont {Hess}},
  \bibinfo {author} {\bibfnamefont {A.}~\bibnamefont {Kyprianidis}}, \bibinfo
  {author} {\bibfnamefont {P.}~\bibnamefont {Becker}}, \bibinfo {author}
  {\bibfnamefont {A.}~\bibnamefont {Lee}}, \bibinfo {author} {\bibfnamefont
  {J.}~\bibnamefont {Smith}}, \bibinfo {author} {\bibfnamefont
  {G.}~\bibnamefont {Pagano}}, \bibinfo {author} {\bibfnamefont {I.-D.}\
  \bibnamefont {Potirniche}}, \bibinfo {author} {\bibfnamefont {A.~C.}\
  \bibnamefont {Potter}}, \bibinfo {author} {\bibfnamefont {A.}~\bibnamefont
  {Vishwanath}}, \bibinfo {author} {\bibfnamefont {N.~Y.}\ \bibnamefont
  {Yao}},\ and\ \bibinfo {author} {\bibfnamefont {C.}~\bibnamefont {Monroe}},\
  }\bibfield  {title} {\bibinfo {title} {Observation of a discrete time
  crystal},\ }\href {https://doi.org/10.1038/nature21413} {\bibfield  {journal}
  {\bibinfo  {journal} {Nature}\ }\textbf {\bibinfo {volume} {543}},\ \bibinfo
  {pages} {217} (\bibinfo {year} {2017})}\BibitemShut {NoStop}%
\bibitem [{\citenamefont {Wadenpfuhl}\ and\ \citenamefont
  {Adams}(2023)}]{wadenpfuhl_emergence_2023}%
  \BibitemOpen
  \bibfield  {author} {\bibinfo {author} {\bibfnamefont {K.}~\bibnamefont
  {Wadenpfuhl}}\ and\ \bibinfo {author} {\bibfnamefont {C.~S.}\ \bibnamefont
  {Adams}},\ }\bibfield  {title} {\bibinfo {title} {Emergence of
  {Synchronization} in a {Driven}-{Dissipative} {Hot} {Rydberg} {Vapor}},\
  }\href {https://doi.org/10.1103/PhysRevLett.131.143002} {\bibfield  {journal}
  {\bibinfo  {journal} {Phys. Rev. Lett.}\ }\textbf {\bibinfo {volume} {131}},\
  \bibinfo {pages} {143002} (\bibinfo {year} {2023})}\BibitemShut {NoStop}%
\bibitem [{\citenamefont {Wu}\ \emph {et~al.}(2023)\citenamefont {Wu},
  \citenamefont {Wang}, \citenamefont {Yang}, \citenamefont {Gao},
  \citenamefont {Liang}, \citenamefont {Tey}, \citenamefont {Li}, \citenamefont
  {Pohl},\ and\ \citenamefont {You}}]{wu_observation_2023}%
  \BibitemOpen
  \bibfield  {author} {\bibinfo {author} {\bibfnamefont {X.}~\bibnamefont
  {Wu}}, \bibinfo {author} {\bibfnamefont {Z.}~\bibnamefont {Wang}}, \bibinfo
  {author} {\bibfnamefont {F.}~\bibnamefont {Yang}}, \bibinfo {author}
  {\bibfnamefont {R.}~\bibnamefont {Gao}}, \bibinfo {author} {\bibfnamefont
  {C.}~\bibnamefont {Liang}}, \bibinfo {author} {\bibfnamefont {M.~K.}\
  \bibnamefont {Tey}}, \bibinfo {author} {\bibfnamefont {X.}~\bibnamefont
  {Li}}, \bibinfo {author} {\bibfnamefont {T.}~\bibnamefont {Pohl}},\ and\
  \bibinfo {author} {\bibfnamefont {L.}~\bibnamefont {You}},\ }\href
  {https://doi.org/10.48550/arXiv.2305.20070} {\bibinfo {title} {Observation of
  a dissipative time crystal in a strongly interacting {Rydberg} gas}}
  (\bibinfo {year} {2023}),\ \Eprint {https://arxiv.org/abs/2305.20070}
  {arXiv:2305.20070 [cond-mat.quant-gas]} \BibitemShut {NoStop}%
\bibitem [{\citenamefont {Mi}\ \emph {et~al.}(2022)\citenamefont {Mi},
  \citenamefont {Ippoliti}, \citenamefont {Quintana}, \citenamefont {Greene},
  \citenamefont {Chen}, \citenamefont {Gross}, \citenamefont {Arute},
  \citenamefont {Arya}, \citenamefont {Atalaya}, \citenamefont {Babbush} \emph
  {et~al.}}]{mi_time-crystalline_2022}%
  \BibitemOpen
  \bibfield  {author} {\bibinfo {author} {\bibfnamefont {X.}~\bibnamefont
  {Mi}}, \bibinfo {author} {\bibfnamefont {M.}~\bibnamefont {Ippoliti}},
  \bibinfo {author} {\bibfnamefont {C.}~\bibnamefont {Quintana}}, \bibinfo
  {author} {\bibfnamefont {A.}~\bibnamefont {Greene}}, \bibinfo {author}
  {\bibfnamefont {Z.}~\bibnamefont {Chen}}, \bibinfo {author} {\bibfnamefont
  {J.}~\bibnamefont {Gross}}, \bibinfo {author} {\bibfnamefont
  {F.}~\bibnamefont {Arute}}, \bibinfo {author} {\bibfnamefont
  {K.}~\bibnamefont {Arya}}, \bibinfo {author} {\bibfnamefont {J.}~\bibnamefont
  {Atalaya}}, \bibinfo {author} {\bibfnamefont {R.}~\bibnamefont {Babbush}},
  \emph {et~al.},\ }\bibfield  {title} {\bibinfo {title} {Time-crystalline
  eigenstate order on a quantum processor},\ }\href
  {https://doi.org/10.1038/s41586-021-04257-w} {\bibfield  {journal} {\bibinfo
  {journal} {Nature}\ }\textbf {\bibinfo {volume} {601}},\ \bibinfo {pages}
  {531} (\bibinfo {year} {2022})}\BibitemShut {NoStop}%
\bibitem [{\citenamefont {Frey}\ and\ \citenamefont
  {Rachel}(2022)}]{frey_realization_2022}%
  \BibitemOpen
  \bibfield  {author} {\bibinfo {author} {\bibfnamefont {P.}~\bibnamefont
  {Frey}}\ and\ \bibinfo {author} {\bibfnamefont {S.}~\bibnamefont {Rachel}},\
  }\bibfield  {title} {\bibinfo {title} {Realization of a discrete time crystal
  on 57 qubits of a quantum computer},\ }\href
  {https://doi.org/10.1126/sciadv.abm7652} {\bibfield  {journal} {\bibinfo
  {journal} {Sci. Adv.}\ }\textbf {\bibinfo {volume} {8}},\ \bibinfo {pages}
  {eabm7652} (\bibinfo {year} {2022})}\BibitemShut {NoStop}%
\bibitem [{\citenamefont {Schindler}\ and\ \citenamefont
  {Sheinfux}(2023)}]{schindler_floquet_2023}%
  \BibitemOpen
  \bibfield  {author} {\bibinfo {author} {\bibfnamefont {S.~T.}\ \bibnamefont
  {Schindler}}\ and\ \bibinfo {author} {\bibfnamefont {H.~H.}\ \bibnamefont
  {Sheinfux}},\ }\href {https://doi.org/10.48550/arXiv.2311.00845} {\bibinfo
  {title} {Floquet engineering with spatially non-uniform driving fields}}
  (\bibinfo {year} {2023}),\ \Eprint {https://arxiv.org/abs/2311.00845}
  {arXiv:2311.00845 [physics.optics]} \BibitemShut {NoStop}%
\bibitem [{\citenamefont {Magnus}(1954)}]{magnus_exponential_1954}%
  \BibitemOpen
  \bibfield  {author} {\bibinfo {author} {\bibfnamefont {W.}~\bibnamefont
  {Magnus}},\ }\bibfield  {title} {\bibinfo {title} {On the exponential
  solution of differential equations for a linear operator},\ }\href
  {https://doi.org/10.1002/cpa.3160070404} {\bibfield  {journal} {\bibinfo
  {journal} {Commun. Pure Appl. Math.}\ }\textbf {\bibinfo {volume} {7}},\
  \bibinfo {pages} {649} (\bibinfo {year} {1954})}\BibitemShut {NoStop}%
\bibitem [{\citenamefont {Blanes}\ \emph {et~al.}(2009)\citenamefont {Blanes},
  \citenamefont {Casas}, \citenamefont {Oteo},\ and\ \citenamefont
  {Ros}}]{blanes_magnus_2009}%
  \BibitemOpen
  \bibfield  {author} {\bibinfo {author} {\bibfnamefont {S.}~\bibnamefont
  {Blanes}}, \bibinfo {author} {\bibfnamefont {F.}~\bibnamefont {Casas}},
  \bibinfo {author} {\bibfnamefont {J.~A.}\ \bibnamefont {Oteo}},\ and\
  \bibinfo {author} {\bibfnamefont {J.}~\bibnamefont {Ros}},\ }\bibfield
  {title} {\bibinfo {title} {The {Magnus} expansion and some of its
  applications},\ }\href {https://doi.org/10.1016/j.physrep.2008.11.001}
  {\bibfield  {journal} {\bibinfo  {journal} {Phys. Rep.}\ }\textbf {\bibinfo
  {volume} {470}},\ \bibinfo {pages} {151} (\bibinfo {year}
  {2009})}\BibitemShut {NoStop}%
\bibitem [{\citenamefont {Yates}\ \emph {et~al.}(2022)\citenamefont {Yates},
  \citenamefont {Abanov},\ and\ \citenamefont {Mitra}}]{yates_long-lived_2022}%
  \BibitemOpen
  \bibfield  {author} {\bibinfo {author} {\bibfnamefont {D.~J.}\ \bibnamefont
  {Yates}}, \bibinfo {author} {\bibfnamefont {A.~G.}\ \bibnamefont {Abanov}},\
  and\ \bibinfo {author} {\bibfnamefont {A.}~\bibnamefont {Mitra}},\ }\bibfield
   {title} {\bibinfo {title} {Long-lived period-doubled edge modes of
  interacting and disorder-free {Floquet} spin chains},\ }\href
  {https://doi.org/10.1038/s42005-022-00818-1} {\bibfield  {journal} {\bibinfo
  {journal} {Commun. Phys.}\ }\textbf {\bibinfo {volume} {5}},\ \bibinfo
  {pages} {1} (\bibinfo {year} {2022})}\BibitemShut {NoStop}%
\bibitem [{\citenamefont {Kitaev}(2001)}]{kitaev_unpaired_2001}%
  \BibitemOpen
  \bibfield  {author} {\bibinfo {author} {\bibfnamefont {A.~Y.}\ \bibnamefont
  {Kitaev}},\ }\bibfield  {title} {\bibinfo {title} {Unpaired {Majorana}
  fermions in quantum wires},\ }\href
  {https://doi.org/10.1070/1063-7869/44/10S/S29} {\bibfield  {journal}
  {\bibinfo  {journal} {Physics-Uspekhi}\ }\textbf {\bibinfo {volume} {44}},\
  \bibinfo {pages} {131} (\bibinfo {year} {2001})}\BibitemShut {NoStop}%
\bibitem [{\citenamefont {Senthil}(2015)}]{senthil_symmetry-protected_2015}%
  \BibitemOpen
  \bibfield  {author} {\bibinfo {author} {\bibfnamefont {T.}~\bibnamefont
  {Senthil}},\ }\bibfield  {title} {\bibinfo {title} {Symmetry-{Protected}
  {Topological} {Phases} of {Quantum} {Matter}},\ }\href
  {https://doi.org/10.1146/annurev-conmatphys-031214-014740} {\bibfield
  {journal} {\bibinfo  {journal} {Annu. Rev. Condens. Matter Phys.}\ }\textbf
  {\bibinfo {volume} {6}},\ \bibinfo {pages} {299} (\bibinfo {year}
  {2015})}\BibitemShut {NoStop}%
\bibitem [{\citenamefont {Bezanson}\ \emph {et~al.}(2017)\citenamefont
  {Bezanson}, \citenamefont {Edelman}, \citenamefont {Karpinski},\ and\
  \citenamefont {Shah}}]{bezansonJuliaFreshApproach2017}%
  \BibitemOpen
  \bibfield  {author} {\bibinfo {author} {\bibfnamefont {J.}~\bibnamefont
  {Bezanson}}, \bibinfo {author} {\bibfnamefont {A.}~\bibnamefont {Edelman}},
  \bibinfo {author} {\bibfnamefont {S.}~\bibnamefont {Karpinski}},\ and\
  \bibinfo {author} {\bibfnamefont {V.~B.}\ \bibnamefont {Shah}},\ }\bibfield
  {title} {\bibinfo {title} {Julia: {{A Fresh Approach}} to {{Numerical
  Computing}}},\ }\href {https://doi.org/10.1137/141000671} {\bibfield
  {journal} {\bibinfo  {journal} {SIAM Rev.}\ }\textbf {\bibinfo {volume}
  {59}},\ \bibinfo {pages} {65} (\bibinfo {year} {2017})}\BibitemShut {NoStop}%
\end{thebibliography}%

\end{document}